\pdfoutput=1
\documentclass[aps,pra,reprint,superscriptaddress,amsmath,amssymb,floatfix]{revtex4-2}

\usepackage{graphicx}
\usepackage{array}
\usepackage{bm}
\usepackage[colorlinks=true,linkcolor=blue,citecolor=blue,urlcolor=blue]{hyperref}

\newcommand{\astar}{a^{*}}
\newcommand{\arot}{a_{\mathrm{rot}}^{*}}
\newcommand{\tQ}{\tau_Q}
\newcommand{\that}{\hat{t}}
\newcommand{\perum}{\,\mu\mathrm{m}^{-1}}
\newcommand{\Dy}{$^{164}$Dy}
\newcommand{\tcell}[2]{\parbox[t]{#1\columnwidth}{\raggedright #2\strut}}

\begin{document}

\title{Spinodal-controlled freeze-out and nucleation at the first-order\\
superfluid--supersolid transition of a dipolar gas}

\author{Wai Kui Wong}
\email{kuchrisuk@gmail.com}
\affiliation{Independent Researcher}

\date{October 8, 2026}

\begin{abstract}
Dipolar quantum gases in elongated traps become supersolid through a
transition that is continuous at intermediate density but first order at low
and high density; at the continuous point, ramps of the scattering length
follow Kibble--Zurek scaling. We study ramps across the first-order
superfluid--supersolid transition of
\Dy{} in an infinite tube using the three-dimensional extended
Gross--Pitaevskii equation with truncated-Wigner noise. With quantum noise
only, the uniform superfluid survives the metastable window and transforms
after its spinodal, where the roton softens. For onsets occurring during the
linear ramp, the delay measured from this spinodal has a fitted exponent
$\zeta = 0.36 \pm 0.02$, close to the continuous-transition result and the
mean-field value $1/3$. The quoted uncertainty combines statistical and
systematic sensitivities. Reverse ramps show a comparable delay beyond the
fold of the crystal branch, producing dynamical hysteresis. Initial thermal
noise with nominal temperatures of 2.5--10\,nK produces crystalline regions
inside the metastable window during holds, with faster nucleation closer to
the spinodal and at higher initial temperature. In ramps at 5\,nK, thermal
fluctuations make the superfluid fraction $f_s$ cross onset thresholds near
$f_s = 1$ early, whereas the bulk of
the crystal still forms after the zero-temperature spinodal with a similar
scaling. Without a heat bath these
effective rates depend on preparation: ramps crossing the window at the same
rate nucleate early in 2 of 32 or 11 of 16 realizations depending on their
starting point. A crossover time inferred from hold rates is therefore an
estimate, not a measured boundary for the ramp protocol. At low density,
slower ramps select larger spacings in a crystal of separated droplets, and
droplets survive far above the transition on reverse ramps. A Maxwell
construction gives a coexistence window $0.037\,a_0$ wide around the
first-order transition point, in which localized states exist; in the
simulated tube they lie above the homogeneous states. At the continuous point
there is no coexistence.
\end{abstract}

\maketitle

\section{Introduction}

A supersolid combines the two defining properties of a crystal and a
superfluid: its density is periodically modulated, yet it flows without
friction and keeps global phase coherence. After decades of searches in solid
helium, supersolid states were realized in dipolar Bose--Einstein condensates
of dysprosium and erbium held in elongated
traps~\cite{Tanzi2019,Bottcher2019,Chomaz2019} (see Ref.~\cite{Chomaz2023} for
a review). In these gases the long-range, anisotropic dipole--dipole
interaction produces a roton minimum in the excitation spectrum. When the
contact scattering length $a_s$ is lowered, the roton energy decreases; once
it reaches zero, the uniform superfluid becomes unstable towards a density
modulation with the roton wavelength. Quantum fluctuations, described by the
Lee--Huang--Yang (LHY) correction, stop the collapse that would otherwise
follow, and a modulated state with a finite superfluid fraction---the
supersolid---forms.

The character of this transition depends on the atom density. For a dipolar
gas in an infinite tube, Blakie \textit{et al.}~\cite{Blakie2020} and Smith,
Baillie and Blakie~\cite{Smith2023} found that the transition is continuous
only in a window of intermediate linear densities (see also
Refs.~\cite{Roccuzzo2019,Biagioni2022,SanchezBaena2024}). At lower and higher
densities it is first order: the modulated state becomes the ground state at a
scattering length $\astar$ before the uniform state becomes unstable at
$\arot$, so that both states are local energy minima over a range of
$a_s$, and hysteresis is possible. This classification compares states of
uniform mean density. Because the atom number is conserved, a long tube can
instead separate into regions of different density, and localized
states---a crystal patch embedded in the superfluid---can then be the ground
state; a supercritical bifurcation at fixed density therefore does not by
itself establish a continuous transition~\cite{Steinberg2025}. Thermal
fluctuations also move the transition: they soften the roton and shift the
phase boundary to larger $a_s$~\cite{SanchezBaena2023,SanchezBaena2024}.

Experiments do not cross a phase transition infinitely slowly. The scattering
length is ramped with a Feshbach resonance over tens to hundreds of
milliseconds, and the gas is at a small but finite temperature. For a
continuous transition, the response to a finite-rate ramp is described by the
Kibble--Zurek (KZ) mechanism~\cite{Kibble1976,Zurek1985}: close to the
critical point the relaxation time diverges, the system can no longer follow
the ramp, and the new order appears only after a delay, the freeze-out time
$\that$, that scales as a power of the ramp time $\tQ$. Kirkby, Salman,
Gasenzer and Chomaz~\cite{Kirkby2025} recently confirmed this picture for the
supersolid transition of \Dy{} at a density where the transition is
continuous, finding $\that \propto \tQ^{\zeta}$ with $\zeta \approx 0.35$,
close to the mean-field value $1/3$. They explicitly restricted their study to
the continuous regime.

\begin{figure*}[t]
\includegraphics[width=\textwidth]{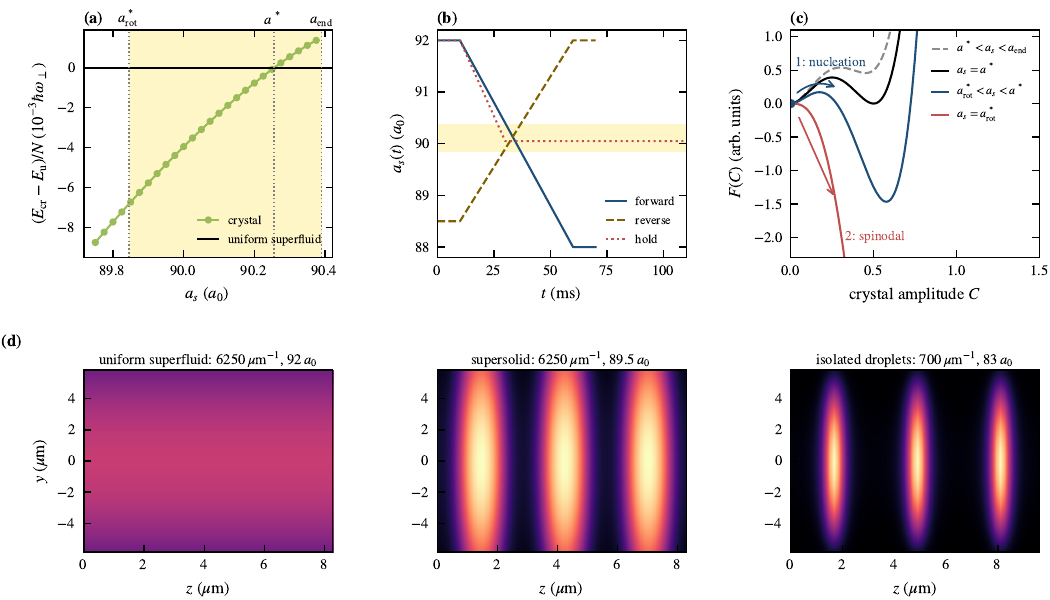}
\caption{(a) Energy per atom of the crystal relative to the uniform superfluid
at $6250\perum$ (static ground states, cell 2.76\,$\mu$m), with the uniform
spinodal $\arot = 89.845\,a_0$, the transition point $\astar = 90.256\,a_0$ and
the end of the crystal branch, $90.39\,a_0$; the metastable window between the
two spinodals is shaded. (b) Ramp protocols: forward ramps (superfluid $\to$
crystal), reverse ramps (crystal $\to$ superfluid) and holds inside the window,
each after 10\,ms of evolution with truncated-Wigner noise. (c) The two routes
to the crystal in a schematic Landau energy $F(C) = rC^2 - C^3 + C^4$ of the
crystal amplitude $C$ ($r = 0$: uniform spinodal; $r = 1/4$: $\astar$;
$r = 9/32$: end of the crystal branch): inside the window the uniform state is
separated from the crystal by a barrier that thermal fluctuations can cross
(route 1); at the spinodal the barrier vanishes (route 2). (d) Column density
$\int dx\,|\Psi|^2$ (dipoles along $y$) of ground states over three cells:
uniform superfluid and supersolid at $6250\perum$ (common color scale), and
isolated droplets at $700\perum$.}
\label{fig:schematic}
\end{figure*}

At a first-order transition the KZ argument is not straightforward. There is
no diverging length or time at the transition point $\astar$, where the two
states have equal energy at equal density: both the uniform and the modulated
states are locally stable there. A
divergence occurs only at the spinodal $\arot$, where the uniform state loses
stability---but the system can avoid reaching it if the new phase nucleates
earlier inside the metastable window, as supercooled water freezes when an ice
nucleus forms. In liquid helium, density-functional studies showed that
cavitation can occur within the metastable region, while the nucleation
barrier vanishes at the spinodal~\cite{Jezek1993,Guilleumas1995}.
Figure~\ref{fig:schematic}(c) sketches the two routes in a
schematic Landau energy of the crystal amplitude: inside the window the
uniform state sits in a local minimum separated by a barrier from the crystal
(route 1: nucleation over the barrier); at the spinodal the barrier vanishes
and the uniform state becomes unstable (route 2). Which of these routes is
taken, and whether a scaling law survives, is the question of this paper.

Power-law scaling across first-order transitions has been reported in other
systems. In a spin-1 condensate driven from the polar to the antiferromagnetic
phase, the onset time of spin excitations scales with the quench rate, which
was described by a generalized KZ mechanism~\cite{Qiu2020}; Gross--Pitaevskii
simulations of a spinor condensate found KZ exponents for the decay of the
metastable phase and for domain formation~\cite{Wheeler2025}; in quantum Ising
chains and classical Potts models the dynamics across first-order transitions
shows spinodal-like scaling, and an emergent critical point near the quantum
spinodal has been proposed~\cite{Pelissetto2025,Pelissetto2026,Zhang2026},
while KZ scaling of defect numbers can only be approximate~\cite{Zhong2025}.
The first-order regime of dipolar supersolids has not been studied
dynamically. It differs from these systems in two respects: the instability at
the spinodal occurs at a finite wavevector, the roton, and the ordered phase is
a crystal whose thermally seeded nucleation competes with the spinodal route
at finite temperature.

We answer this question with simulations of the full three-dimensional
extended Gross--Pitaevskii equation (eGPE) for \Dy{} in a tube, at linear
densities of 700, 2500 and $6250\perum$, with truncated-Wigner noise
representing quantum or thermal fluctuations. Our main findings are:
\begin{enumerate}
\item At zero temperature (quantum noise only) the uniform superfluid survives
  through the whole metastable window, and the crystal forms only after the
  roton softens. Measured from the spinodal, the freeze-out time follows a
  KZ-like power law with an exponent close to that of the continuous
  transition, $\zeta \approx 1/3$, over the tested ramp window; post-ramp
  onsets are excluded from the linear-ramp fit. Measured from $\astar$, the
  forward onset curve is not a single power law.
\item On reverse ramps the crystal persists up to the end of its own branch and
  melts after a delay with a comparable power-law scaling. The two spinodals
  bound a dynamical hysteresis loop that widens as the ramp becomes faster.
\item With thermal noise initialized at nominal temperatures of
  2.5--10\,nK, crystalline regions form inside the metastable window.
  Nucleation is faster close to the spinodal and at higher initial
  temperature; near $\astar$ fewer realizations nucleate, but a nucleated
  crystal still fills the tube. In ramps at 5\,nK the superfluid fraction
  starts to fall earlier and falls more gradually, so that onset times depend
  on the threshold; for $f_s < 0.5$ the onsets still follow the
  zero-temperature spinodal, with an exponent similar to the quantum-noise one.
  Hold rates suggest a
  crossover to nucleation for slower ramps, but its location is not measured
  and depends on the
  preparation through the population of the soft roton mode, which is not
  re-thermalized in truncated-Wigner dynamics.
\item At low density, where the modulated state consists of separated
  droplets, slow ramps select larger droplet spacings, and droplets
  survive far above the transition on reverse ramps.
\item In the static phase diagram, a Maxwell construction gives a window
  $0.037\,a_0$ wide around $\astar$ in which states of different density
  coexist at $6250\perum$, and no coexistence at $2500\perum$, where the
  transition is therefore continuous also in this sense. Localized states
  exist in the simulated tube but lie above the homogeneous states: their
  interfaces cost more than phase separation gains in tubes shorter than
  about $250\,\mu$m.
\end{enumerate}

Section~\ref{sec:model} describes the model, the static phase diagram and the
numerical method. Section~\ref{sec:results} presents the results, and
Sec.~\ref{sec:discussion} discusses their origin, the role of temperature and
the relevance for experiments. Appendices compare our full 3D model with the
quasi-one-dimensional reduced model used in earlier work, list convergence
tests, give the noise sampling and the error budget, describe the
coexistence analysis and the localized states, and analyze thermal effects on
the onset.

\section{Model and methods}\label{sec:model}

\subsection{Extended Gross--Pitaevskii equation}

We describe the condensate wavefunction $\Psi(\mathbf{r},t)$, normalized to the
atom number $N$, by the eGPE
\begin{equation}
i\hbar\,\partial_t\Psi = \Big[-\frac{\hbar^2\nabla^2}{2m} + V(\mathbf{r})
+ g|\Psi|^2 + \Phi_{\mathrm{dd}}(\mathbf{r},t) + \gamma_{\mathrm{QF}}|\Psi|^3\Big]\Psi ,
\label{eq:egpe}
\end{equation}
with the contact coupling $g = 4\pi\hbar^2 a_s/m$, the dipolar potential
$\Phi_{\mathrm{dd}} = \int d^3r'\, U_{\mathrm{dd}}(\mathbf{r}-\mathbf{r}')|\Psi(\mathbf{r}')|^2$,
$U_{\mathrm{dd}}(\mathbf{r}) = (\mu_0\mu_m^2/4\pi)(1-3\cos^2\theta)/r^3$, and the
LHY coefficient $\gamma_{\mathrm{QF}} = \frac{32}{3} g \sqrt{a_s^3/\pi}\,
Q_5(\epsilon_{\mathrm{dd}})$, where $\epsilon_{\mathrm{dd}} = a_{\mathrm{dd}}/a_s$
and $Q_5$ is the usual auxiliary function~\cite{Lima2011,Wachtler2016},
evaluated exactly. We use \Dy{} ($a_{\mathrm{dd}} = 130.8\,a_0$) in a tube: the
gas is trapped harmonically in the transverse plane,
$V = m\omega_\perp^2(x^2+y^2)/2$ with $\omega_\perp = 2\pi\times150$\,Hz, and
free and periodic along the tube axis $z$. The dipoles point along a
transverse direction ($y$). Lengths are given in units of
$l = \sqrt{\hbar/m\omega_\perp} = 0.641\,\mu$m and times in $1/\omega_\perp$, but
we quote times in ms and densities as atoms per $\mu$m.

\subsection{Static phase diagram}\label{sec:static}

The ramps are interpreted with three characteristic scattering lengths, which
we compute on the same numerical grid as the dynamics
[Fig.~\ref{fig:schematic}(a)]:
\begin{itemize}
\item the \textit{roton instability} $\arot$, where the lowest Bogoliubov
  frequency of the uniform tube vanishes; it is the spinodal of the uniform
  state. We obtain it from the Bogoliubov--de Gennes equations of the
  $z$-uniform ground state, using that $\omega_{\mathrm{rot}}^2$ is linear in
  $a_s$ near the instability;
\item the \textit{equal-density transition point} $\astar$, where the energy per
  atom of the modulated ground state, minimized over the cell length, equals
  that of the uniform state at the same mean density (in the thermodynamic
  limit a narrow coexistence window surrounds it, see below); for brevity we
  call it the transition point;
\item the \textit{end of the crystal branch}, the largest $a_s$ at which a
  modulated stationary state still exists when it is followed upwards in
  $a_s$ by continuation; it is the spinodal of the crystal.
\end{itemize}

At $2500\perum$ the transition is continuous, $\astar = \arot = 92.314\,a_0$.
At $6250\perum$ it is first order: $\arot = 89.845\,a_0 < \astar = 90.256\,a_0$,
and the crystal branch ends at $90.39\,a_0$. Between the two spinodals
(89.85--90.39\,$a_0$) the uniform state, the crystal, or both are locally
stable---the metastable window of Fig.~\ref{fig:schematic}(a). At $700\perum$
the transition is also first order ($\arot = 84.506\,a_0$,
$\astar = 85.28\,a_0$) and leads directly from the uniform superfluid to a
crystal of isolated droplets whose optimal spacing grows without bound near
$\astar$. Figure~\ref{fig:schematic}(d) shows the column densities of the
three kinds of state: the uniform superfluid; the supersolid at $6250\perum$,
in which elongated droplets are connected by a superfluid background (Leggett
fraction $f_s = 0.47$ at $89.5\,a_0$); and the crystal of isolated droplets at
$700\perum$ ($f_s = 0.16$ at $83\,a_0$).

These values agree with the full-eGPE literature where available
($\arot = 92.32\,a_0$ at $2500\perum$~\cite{Blakie2023}). The reduced quasi-1D
model used in earlier studies gives values 1--4.5\,$a_0$ lower and a
first-order window about 2.5 times wider than the full 3D theory
(Appendix~\ref{app:quasi1d}). First-order dynamics is particularly sensitive
to such errors in the static phase boundaries: the freeze-out is measured from
the spinodal, and shifting the reference by only $0.03\,a_0$ changes the fitted
exponent by about 0.02 (Sec.~\ref{sec:zeroT}). Quantitative statements about
the first-order regime therefore require the full 3D model used here, with the
transition points computed on the grid of the dynamics.

\textit{Coexistence.} The transition point $\astar$ compares states of equal
mean density. With the atom number fixed, a long tube can lower its energy
further by separating into a crystal at density $n_M$ and a uniform
superfluid at $n_U$ with equal chemical potential and pressure~\cite{Steinberg2025}.
A Maxwell construction on the static states (Appendix~\ref{app:coexistence})
gives, at $6250\perum$, coexisting densities that differ by $53\perum$
(0.85\%), the crystal being less dense; at fixed $n = 6250\perum$ phase
separation is favorable in the thermodynamic limit for
$90.237 < a_s < 90.274\,a_0$, a window $0.037\,a_0$ wide around $\astar$ and 15
times narrower than the metastable window. At $2500\perum$ the chemical
potential of the crystal branch increases with density near its onset and
there is no coexistence, so the transition there is continuous also in this
sense. Localized states---a crystal patch at $n_M$ embedded in superfluid at
$n_U$---exist as stationary states in the simulated tube, but they lie above
the homogeneous states: each of their two interfaces costs about
$48\,\hbar\omega_\perp$, and the bulk-plus-interface estimate predicts that
phase separation becomes favorable only in tubes longer than about
$250\,\mu$m near $\astar$, three times the tube simulated here. The dynamics below is referred
to the two spinodals, which are properties of the uniform and modulated
branches and do not depend on coexistence.

\subsection{Ramps and noise}

Each simulation starts from the uniform ground state at an initial scattering
length $a_i$ above the transition (or from the crystal ground state below it,
for reverse ramps). We add noise, let the gas evolve for 10\,ms at $a_i$, and
then change the scattering length linearly to a final value $a_f$ over the
ramp time $\tQ$, $a_s(t) = a_i + (a_f-a_i)(t-10\,\mathrm{ms})/\tQ$, followed by a
hold at $a_f$ [Fig.~\ref{fig:schematic}(b)]. Forward ramps go from 92 to
$88\,a_0$ at $6250\perum$, from 96 to $88\,a_0$ at $2500\perum$, and from 87 to
$83\,a_0$ at $700\perum$; $\tQ$ ranges from 10 to 400\,ms. The tube contains 32
cells of the ground-state crystal (2.76, 2.83 and 3.2\,$\mu$m at 6250, 2500 and
$700\perum$, i.e., 88--102\,$\mu$m in total).

Fluctuations are included in the truncated-Wigner
approximation~\cite{Steel1998,Sinatra2002,Blakie2008}: every mode of the
initial state receives a complex Gaussian amplitude with mean occupation
$n_{\mathrm{BE}}(\varepsilon)+1/2$. For ``quantum noise'' only the half quantum is
added, in plane waves along the tube with wavenumbers below the inverse healing
length; this adds 0.01--0.06\% of the atoms. For thermal noise at temperature
$T$ we populate plane waves times transverse oscillator states with energies
below $2k_BT$ (Appendix~\ref{app:noise}); at 2.5, 5 and 10\,nK this adds 0.19,
0.40 and 0.84\% of the atoms. We use 8--24 realizations per parameter set.

\subsection{Observables}

The order parameter is the superfluid fraction of the line density
$n(z) = \int dx\,dy\,|\Psi|^2$, estimated by Leggett's bound
$f_s = L^2/[N\int_0^L dz/n(z)]$, which equals one for a uniform gas,
approaches zero for isolated droplets, and is an upper bound on the superfluid
fraction along the tube that is close to it for a modulation along one
direction. For evolving fields this density functional is an onset
diagnostic; it does not independently establish phase coherence or a
thermodynamic superfluid response. The \textit{freeze-out time} $\that$ is
the time, measured from the
moment the ramp crosses a reference scattering length, at which $f_s$ of a
realization first drops below 0.98; we report the median over realizations,
which is less sensitive than the mean to a single realization that transforms
early. The reference is either the spinodal $\arot$ or the transition point
$\astar$. We also count droplets---local maxima of $n(z)$ above 1.2 (or, as a
stricter criterion, 1.5) times the mean density.

\subsection{Numerics}

Equation~(\ref{eq:egpe}) is integrated with a Fourier split-step (Strang)
method on a $96\times96\times N_z$ grid (spacing $0.25\,l = 0.16\,\mu$m,
transverse box $24\,l$), with the dipolar interaction cut off at half the
transverse box to suppress periodic images~\cite{Smith2023,Ronen2006,Lu2010}.
The production time step is $\Delta t = 0.00942478/\omega_\perp =
10\,\mu$s, and onset diagnostics are recorded every 0.1\,ms.
Single-precision (complex64) arithmetic gives freeze-out times identical to
double precision in every realization tested; reducing the grid spacing to
$0.1875\,l$ changes $\that$ by at most 1.6\%, enlarging the transverse box to
$32\,l$ by 2\%, and doubling the tube length by less than 1\%
(Appendix~\ref{app:convergence}). The code was validated against analytic
results and published supersolid results, including the freeze-out exponent
of Ref.~\cite{Kirkby2025} in its own reduced model ($\zeta = 0.355\pm0.004$
versus 0.346--0.352)~\cite{WongDipGPE}.

\section{Results}\label{sec:results}

\subsection{Zero temperature: freeze-out from the spinodal}\label{sec:zeroT}

\begin{figure*}[t]
\includegraphics[width=\textwidth]{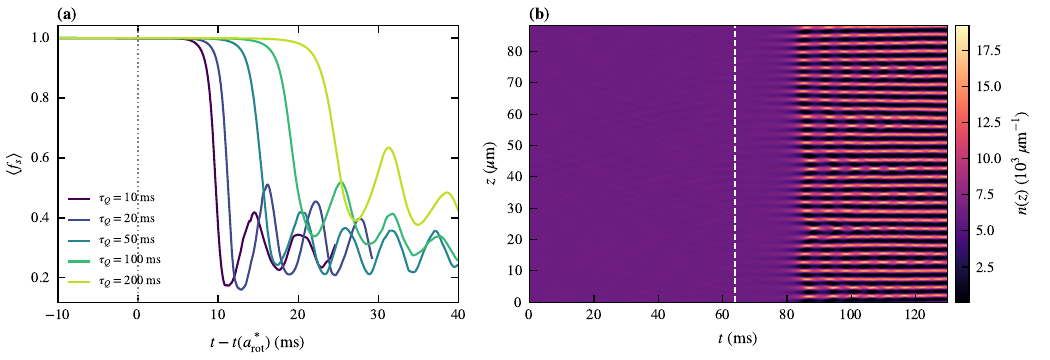}
\caption{Forward ramps at $6250\perum$ with quantum noise. (a) Mean Leggett
superfluid fraction versus time after the ramp crosses the spinodal $\arot$,
for $\tQ = 10$--200\,ms (16 realizations each). The gas stays uniform through
the metastable window and transforms after a delay that grows with $\tQ$.
(b) Line density $n(z,t)$ of one realization for $\tQ = 100$\,ms; the dashed
line marks the spinodal crossing. The modulation forms with independent phases
in different parts of the tube.}
\label{fig:dynamics}
\end{figure*}

We first consider forward ramps with quantum noise only.
Figure~\ref{fig:dynamics}(a) shows the mean superfluid fraction at
$6250\perum$ as a function of the time elapsed since the ramp crossed the
spinodal $\arot$. For every ramp time $f_s$ stays at one well after $\arot$ has
been passed, then drops within a few milliseconds to 0.2--0.4, and oscillates
as the newly formed crystal rings. The drop happens later, in time after the
spinodal, for slower ramps. Before the drop the density is uniform to within
the noise; afterwards a modulation with the roton wavelength has formed,
initially with independent phases in different parts of the tube
[Fig.~\ref{fig:dynamics}(b)]: domains separated by phase defects, as expected
for KZ dynamics.

The key observation is what does \textit{not} happen. Between
$\astar = 90.256\,a_0$ and $\arot = 89.845\,a_0$ the crystal is already the
lower-energy state, so the uniform superfluid could in principle transform
there. It does not: the gas remains uniform throughout the metastable window
and transforms only after the roton instability. Holding the scattering length
at $90.05\,a_0$, in the middle of the window, confirms this: with quantum noise
the gas stays uniform for 150\,ms (Sec.~\ref{sec:nucleation}).

\begin{figure*}[t]
\includegraphics[width=\textwidth]{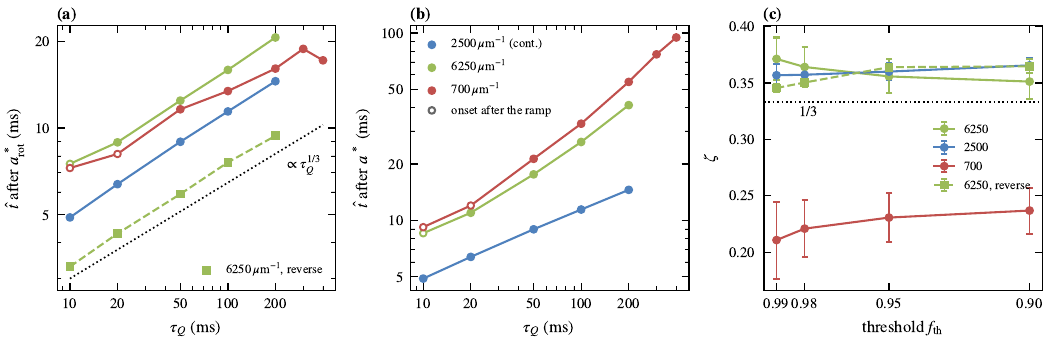}
\caption{Freeze-out times (median over 16 realizations, $f_s < 0.98$).
(a) Measured from the spinodal: $6250\perum$ (first order), $2500\perum$
(continuous), $700\perum$ (first order, droplet crystal) and reverse ramps at
$6250\perum$ measured from the end of the crystal branch; the dotted line is
$\tQ^{1/3}$. Open symbols mark onsets during the final hold, which are
excluded from the linear-ramp fits; squares and dashed lines mark reverse
ramps.
(b) The forward data measured from the equal-density transition
point $\astar$: the first-order data are curved. (c) Fitted exponent $\zeta$
versus the freeze-out threshold $f_{\mathrm{th}}$, with bootstrap error bars.
Each threshold curve uses a common ramp window in which all four threshold
medians precede the ramp end: 50--200\,ms (6250 forward), 20--200\,ms
(2500), 50--400\,ms (700), and 10--200\,ms (6250 reverse).
The low-density slope is descriptive because its onset curve is not a
single power law.}
\label{fig:freezeout}
\end{figure*}

The freeze-out times are collected in Fig.~\ref{fig:freezeout}. At
$6250\perum$, the $\tQ=10$\,ms ensemble crosses the onset threshold during
the final hold: the median onset is at 22.9\,ms, after the ramp ends at
20\,ms. We retain this point as a protocol-dependent onset measurement but
exclude it from the linear-ramp scaling fit. Fitting the remaining
$\tQ=20$--200\,ms points gives
\begin{equation}
\zeta = 0.363 \pm 0.021 \quad \text{at } 6250\perum \text{ (first order)},
\end{equation}
compared with
\begin{equation}
\zeta = 0.364 \pm 0.019 \quad \text{at } 2500\perum \text{ (continuous)},
\end{equation}
where the continuous-transition fit uses $\tQ=10$--200\,ms. Each point has
16 realizations. Including the post-ramp first-order point would give
$\zeta=0.342$.
Both fitted exponents are close to the
mean-field KZ value $\zeta = 1/3$ and to the value 0.35 found by Kirkby
\textit{et al.} at the continuous point. Measured instead from the
equal-density transition point $\astar$, the $6250\perum$ data are not a power
law: the log--log plot is curved, with a local slope that grows from about 0.4
to 0.7 [Fig.~\ref{fig:freezeout}(b)]. The spinodal, not $\astar$, is the point
that controls the freeze-out.

This has a simple explanation. The KZ freeze-out exponent is
$\zeta = z\nu/(1+z\nu)$, where $\nu$ and $z$ describe how the relaxation time
grows near the instability. In mean-field theory the roton gap closes as
$\varepsilon_{\mathrm{rot}} \propto (a_s-\arot)^{1/2}$, which gives $z\nu = 1/2$
and $\zeta = 1/3$. At the continuous transition the roton instability
coincides with $\astar$; at the first-order transition there is no
criticality at $\astar$, but the metastable uniform branch, protected by the
absence of nucleation, terminates at a spinodal with the same kind of soft
roton mode, and the freeze-out near this spinodal follows the same scaling.
The scaling is therefore spinodal-controlled and KZ-like: it is not a property
of the transition point $\astar$ (or of the narrow coexistence window around
it) but of the local instability that ends the metastable branch. The two fitted exponents are numerically equal within
the reported sensitivity estimates. These estimates combine in quadrature
the bootstrap standard deviation, half the threshold-dependent slope range,
the effect of a $\pm0.01\,a_0$ reference shift, the shift between the primary
fit and the common threshold window, and a spatial-sensitivity allowance of
0.015. Threshold tests use a fixed cohort of ramp times for which all
threshold medians precede the ramp end. The estimates are not confidence
intervals and do not establish an asymptotic exponent from the limited ramp
range. Appendix~\ref{app:errors} gives the components.

A practical consequence is that the exponent is very sensitive to the
reference point. Because the freeze-out time is short compared with the time
the ramp spends between $a_i$ and $a_f$, shifting the reference by $\delta a$
changes $\that$ by $\tQ\,\delta a/(a_i-a_f)$, an amount that grows with $\tQ$.
A shift of $+0.03\,a_0$ changes the first-order $\zeta$ by
about 0.023. We
therefore compute $\arot$ on the same numerical grid as the dynamics
(Sec.~\ref{sec:static}); with the value of a slightly different grid, the
apparent exponent would change by more than its statistical error.

\subsection{Reverse ramps and dynamical hysteresis}\label{sec:reverse}

A first-order transition should also show the reverse effect: a crystal that
survives when the scattering length is raised above $\astar$. We start from the
crystal ground state at $88.5\,a_0$ (32 cells of 2.76\,$\mu$m, line contrast
0.92) and ramp $a_s$ up to $92\,a_0$. The superfluid fraction rises gradually
at first, because the superfluid fraction of the supersolid branch itself
grows with $a_s$, and then jumps to one when the crystal melts. Melting does
not happen at $\astar = 90.256\,a_0$: the crystal persists up to and beyond the
end of its branch, $90.39\,a_0$, and melts after a delay that again follows a
power law in $\tQ$. Measured from the branch end, the median melting delays
are 3.3, 4.3, 5.9, 7.6 and 9.4\,ms for $\tQ = 10$, 20, 50, 100 and 200\,ms,
giving $\zeta = 0.351 \pm 0.025$, numerically comparable to the forward
exponent. (The branch end from the Bogoliubov soft mode below, $90.38\,a_0$,
differs from the ground-state value by less than the $\pm0.01\,a_0$ reference
uncertainty included in the error.)

\begin{figure*}[t]
\includegraphics[width=\textwidth]{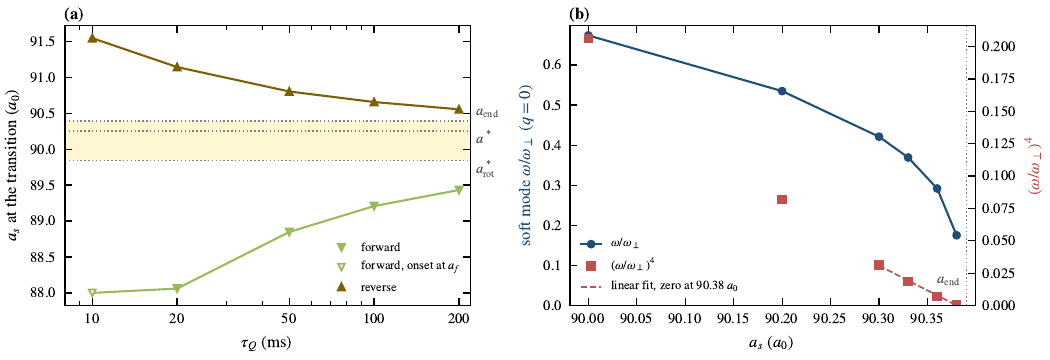}
\caption{(a) Dynamical hysteresis at $6250\perum$: scattering length at which
the transition occurs (median $f_s$ crossing) on forward ramps (superfluid
$\to$ crystal) and reverse ramps (crystal $\to$ superfluid) versus $\tQ$, with
the two spinodals and $\astar$. The open triangle is the $10$\,ms forward
onset during the final hold at $88\,a_0$; all other onset locations are
within their linear ramps. (b) Lowest Bogoliubov frequency of the crystal
at quasi-momentum $q = 0$ (soft amplitude mode) near the end of its branch, and
$\omega^4$ (right axis, squares) with a linear fit that vanishes at $90.38\,a_0$, the
square-root law of a fold.}
\label{fig:hysteresis}
\end{figure*}

Figure~\ref{fig:hysteresis}(a) combines both directions. On forward ramps the
transition ($f_s < 0.98$) occurs at $a_s = 88.00$, 88.06, 88.85, 89.21 and
$89.43\,a_0$ for $\tQ = 10$--200\,ms, i.e., below the spinodal $\arot$ by an
amount that shrinks with $\tQ$; on reverse ramps it occurs at 91.55, 91.14,
90.80, 90.66 and $90.55\,a_0$ for $\tQ = 10$--200\,ms, above the branch end.
The $10$\,ms forward point is at the held final value $88\,a_0$; continuing
the linear ramp formula beyond its endpoint would incorrectly place it at
$86.84\,a_0$. The result is a dynamical hysteresis loop. When both onsets
occur during their ramps, its width is the static spinodal window
($0.55\,a_0$) plus $r_{\mathrm{f}}\that_{\mathrm{f}}+
r_{\mathrm{r}}\that_{\mathrm{r}}$, where the positive ramp rates are
$r_{\mathrm{f}}=4a_0/\tQ$ and $r_{\mathrm{r}}=3.5a_0/\tQ$.
If each delay scales as $\tQ^{\zeta}$, its contribution shrinks as
$\tQ^{\zeta-1}$, close to the $\mathrm{rate}^{2/3}$ dynamical-hysteresis
law of mean-field bistable systems~\cite{Jung1990}. This conversion does not
apply to the forward $10$\,ms hold point. Using the actual scattering length,
the observed width decreases from $3.55\,a_0$ at $10$\,ms to
$1.12\,a_0$ at $200$\,ms; it remains larger than the static window over
the measured range.

The melting exponent is not explained by the argument of
Sec.~\ref{sec:zeroT}. The end of the crystal branch is a saddle-node (fold) of
the crystal's energy landscape rather than a roton instability of a uniform
state. Bogoliubov spectra of the crystal confirm this
[Fig.~\ref{fig:hysteresis}(b)]: the softening mode is the $q = 0$ amplitude
mode of the crystal, its frequency falls from $0.67\,\omega_\perp$ at
$90.0\,a_0$ to $0.18\,\omega_\perp$ at $90.38\,a_0$, $\omega^4$ is linear in
$a_s$ with a zero at $90.38\,a_0$---the square-root law
$\omega^2 \propto (a_{\mathrm{end}}-a_s)^{1/2}$ of a fold, at the branch end found
from the ground states---and no mode at finite quasi-momentum softens first.
For a single mode, the slow passage through a fold delays the escape by
$\that \propto \tQ^{1/5}$ if the dynamics is conservative (equivalently, the KZ
estimate with a gap $\propto \varepsilon^{1/4}$), and by $\that \propto \tQ^{1/3}$
if it is overdamped, the mean-field law behind dynamical
hysteresis~\cite{Jung1990}. The measured melting exponent, $0.351\pm0.025$, is
far from $1/5$ and close to $1/3$, so the melting does not follow the undamped
soft mode. A plausible reason is that the soft mode decays into the gapless
phonon branches once the crystal is driven past the fold, which acts as an
effective damping; we have not tested this.

Hysteresis in the supersolid transition is directly accessible
experimentally: the contrast or the number of droplets can be imaged in situ
during forward and reverse ramps, and the predicted widening of the loop with
ramp rate is a test of the spinodal picture.

\subsection{Thermally seeded nucleation}\label{sec:nucleation}

Quantum noise alone never nucleated the crystal inside the metastable window.
Real gases are at a finite temperature, and thermal fluctuations could provide
the seeds that are missing at zero temperature. We test this directly by
ramping to a scattering length inside the window and holding it there. At
$90.05\,a_0$ the uniform state is linearly stable---its lowest Bogoliubov
frequency is positive ($\omega_{\mathrm{rot}} \approx 0.19\,\omega_\perp$)---so a
crystal can only appear by crossing an energy barrier.

In the language of this section, ``nucleation'' means the formation of
crystalline regions from the initial thermal (classical-field) fluctuations
within the truncated-Wigner dynamics, which has no external heat bath; we do
not claim an activation law $\Gamma \propto \exp(-\Delta F/k_BT)$, which would
require nucleation rates over a wider range of temperatures.

\begin{figure*}[t]
\includegraphics[width=\textwidth]{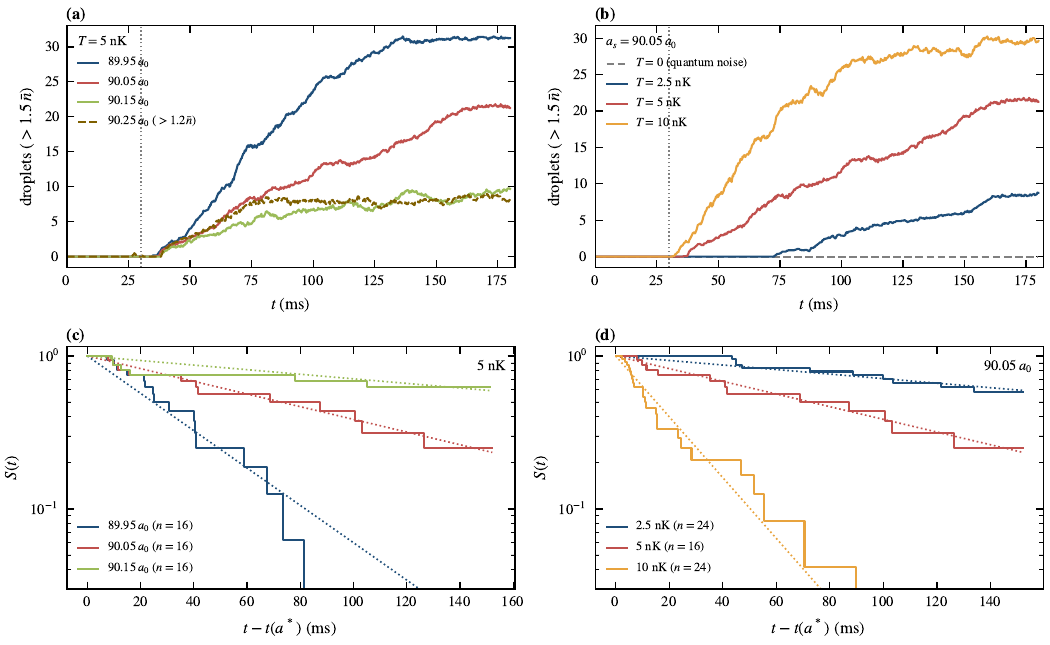}
\caption{Holds inside the metastable window at $6250\perum$ after a 20\,ms ramp
from $92\,a_0$ (the ramp ends at 30\,ms). (a) Mean number of droplets denser
than 1.5 times the mean density at 5\,nK for four scattering lengths (16
realizations; $90.25\,a_0$: 8 realizations, looser 1.2 criterion, dashed);
single realizations stay uniform or crystallize completely.
(b) Temperature dependence at $90.05\,a_0$; with quantum noise only no droplets
form. (c), (d) Fraction $S(t)$ of realizations without a deep droplet versus
time after the ramp crosses $\astar$, at 5\,nK for three scattering lengths (c)
and at $90.05\,a_0$ for three temperatures (d); dotted lines are the
maximum-likelihood exponentials $\exp(-\widehat{\Gamma} L t)$ of Table~\ref{tab:gamma}.}
\label{fig:nucleation}
\end{figure*}

Figure~\ref{fig:nucleation}(a) shows the mean number of droplets denser than
1.5 times the mean density during holds at four scattering lengths across the
window, with thermal noise at 5\,nK (16 realizations; 8 with the looser 1.2
criterion at $90.25\,a_0$). In the three holds with 16 realizations
(89.95--90.15\,$a_0$) the earliest deep droplets appear 5--8\,ms after the
hold begins (7.6--9.3\,ms after the ramp crosses $\astar$, the reference of
Table~\ref{tab:gamma}). After 150\,ms the mean count falls from 31
droplets (a complete crystal) at $89.95\,a_0$ to 21 and 10 at 90.05 and
$90.15\,a_0$ and about 8 at $90.25\,a_0$. This decrease reflects how many
realizations nucleate, not slower growth: individual realizations either stay
uniform or crystallize completely (16/16, 12/16, 6/16 and 2/8 nucleate at the
four scattering lengths), and once the first droplet (1.2 criterion) has
appeared, the count reaches 25 within a median of 32--50\,ms at all four; the
few realizations that are only partly crystalline at the end of the run
nucleated late and are still growing. No realization stops at a crystal patch
beside the superfluid, although $90.25\,a_0$ lies inside the coexistence
window of Sec.~\ref{sec:static} (also at the 0.4\% higher density of the
thermal holds); in the 88\,$\mu$m tube the computed localized states lie
above the homogeneous states (Appendix~\ref{app:coexistence}). With quantum
noise the same hold at $90.05\,a_0$ produces no droplets at all.

Figure~\ref{fig:nucleation}(b) shows the dependence on temperature at
$90.05\,a_0$. Nucleation starts later, and fewer realizations nucleate, as
the temperature is lowered: after the 150\,ms hold there are on average 30, 21
and 9 droplets at 10, 5 and 2.5\,nK, and none at zero temperature. No
temperature dependence is resolved in the time from the first droplet to 25
droplets (median 40--42\,ms, among the 9, 16 and 24 realizations that reach 25
droplets at 2.5, 5 and 10\,nK); at 10\,nK the first droplets form even during
the ramp into the window. The looser droplet
criterion and the superfluid fraction of the density averaged over 1\,ms show
the same behavior, so the droplets are real crystalline regions and not
thermal density fluctuations of the soft roton mode, which can lower the
instantaneous superfluid fraction near the spinodal.

\textit{Nucleation rate.} The hold simulations give a nucleation rate per unit
length and time. For each realization $i$ we record the waiting time $t_i$
from the moment the ramp crosses $\astar$ until the first deep droplet (denser
than 1.5 times the mean) appears; for realizations without a droplet at the end
of the run, $t_i$ is the right-censored observation time. If nuclei form
independently along the tube, the waiting times are exponentially distributed
with rate $\Gamma L$, and the maximum-likelihood estimate is
\begin{equation}
\widehat{\Gamma}=\frac{N_{\mathrm{event}}}{L\sum_i t_i},
\end{equation}
where $N_{\mathrm{event}}$ is the number of nucleated realizations and the sum
runs over all realizations, including the censored ones. The relative
statistical error is $1/\sqrt{N_{\mathrm{event}}}$. The survival curves $S(t)$
[Figs.~\ref{fig:nucleation}(c) and \ref{fig:nucleation}(d)] are consistent with exponentials after a short
delay, which we interpret as the time a nucleus needs to grow to a detectable
droplet. $\Gamma$ is therefore an effective rate of detectable nuclei
(Table~\ref{tab:gamma}).

\begin{table}[t]
\caption{Nucleation rate per unit length and time from holds at $6250\perum$
(maximum-likelihood estimate with right censoring). The events column counts
nucleated realizations out of the total. The delay is the earliest waiting
time, $\min_i t_i$ over the nucleated realizations, the maximum-likelihood
shift of a delayed exponential; $\widehat{\Gamma}_{\mathrm d}$ is the
corresponding rate. Both rates are in $\mu\mathrm{m}^{-1}\mathrm{s}^{-1}$.}
\label{tab:gamma}
\begin{ruledtabular}
\small
\setlength{\tabcolsep}{2pt}
\begin{tabular}{@{}cccccc@{}}
$a_s/a_0$ & $T$ (nK) & events & $\widehat{\Gamma}$ & delay (ms) & $\widehat{\Gamma}_{\mathrm d}$ \\
\hline
89.95 & 5   & 16/16 & $0.32\pm0.08$   & 7.6  & $0.41\pm0.10$ \\
90.05 & 5   & 12/16 & $0.11\pm0.03$   & 7.9  & $0.12\pm0.04$ \\
90.15 & 5   & 6/16  & $0.039\pm0.016$ & 9.3  & $0.043\pm0.017$ \\
90.05 & 2.5 & 10/24 & $0.039\pm0.012$ & 43.5 & $0.060\pm0.019$ \\
90.05 & 10  & 24/24 & $0.52\pm0.11$   & 2.3  & $0.58\pm0.12$ \\
\end{tabular}
\end{ruledtabular}
\end{table}

The rate increases by a factor of about eight across $0.2\,a_0$ of the window,
from near $\astar$ to near the spinodal, and by a factor of about thirteen from
2.5 to 10\,nK. Over this small range of temperatures the dependence is closer
to a power law ($\Gamma \propto T^{\approx 2}$) than to a constant activation
energy, consistent with our decision not to interpret $\Gamma$ in terms of an
Arrhenius law. For the 88\,$\mu$m tube simulated here,
$\Gamma = 0.1\,\mu\mathrm{m}^{-1}\mathrm{s}^{-1}$ corresponds to about one
nucleation event every 110\,ms.

\textit{Does the hold rate predict the ramps?} If nuclei formed at the rate
$\Gamma(a_s)$ measured in the holds, a ramp crossing the window at rate
$r = 4\,a_0/\tQ$ would produce on average $m = L\int\Gamma\,da/r$ nuclei before
reaching the spinodal, and would nucleate there with probability
$1-e^{-m}$: 0.07, 0.13, 0.25 and 0.43 at 5\,nK for $\tQ = 50$, 100, 200 and
400\,ms, with a hold-rate estimate of the probability-one-half crossover at
$\tQ \approx 0.5$\,s. This extrapolation is not a direct measurement of a
crossover in the ramps.
We tested this directly with 16-realization ramps at the rate of
$\tQ = 200$\,ms. Starting, as all production ramps, at $92\,a_0$, only 2 of 32
realizations formed a droplet before the spinodal (predicted: 8; the binomial
probability of two or fewer is below 1\%). Starting instead at $90.5\,a_0$,
just above the window, and crossing it at exactly the same rate, 11 of 16 did.

The difference lies in the soft roton mode. The thermal noise is sampled at
the initial scattering length in the free-particle basis, and the
truncated-Wigner evolution has no bath to thermalize it afterwards. The noise
projects onto the roton with a weight that grows as the roton frequency falls,
so noise injected near the window loads the roton more; during a slow ramp
from far above, the roton population instead follows adiabatically and stays
below the thermal value $k_BT/\hbar\omega_{\mathrm{rot}}$, which grows towards
the spinodal. The roton-band power of the density fluctuations at $\astar$
confirms this: 0.04 for the slow ramp from $92\,a_0$, 0.05 after the fast 20\,ms
ramp that starts the holds, and 0.1--0.2 for ramps from $90.5\,a_0$ (arbitrary
units). The nucleation rate in our simulations is therefore a property of the
protocol as well as of $(a_s, T)$: the hold rates overpredict early nucleation
for the tested slow ramp from $92\,a_0$, while preparation near the window
leads to much earlier nucleation. The recorded ramp ensembles do not locate
the crossover or establish $0.5$\,s as a lower bound for it. A thermalizing
cloud could change the roton population during a ramp; its effect is not
represented by the present preparation protocols.

Thermal noise also changes the onset of the ramps, in a way that depends on
the threshold (Table~\ref{tab:thermal}). With quantum noise the superfluid
fraction falls from 0.98 to 0.5 within about 5\,ms; with thermal noise at
5\,nK it starts to fall earlier and falls more gradually. For $\tQ = 200$\,ms
the thermal onsets are earlier by 11.3, 7.4 and 0.9\,ms for the thresholds
0.98, 0.7 and 0.5; for $\tQ = 50$\,ms by 1.9, 1.6 and 1.2\,ms. The onset curves
are therefore not shifted rigidly. The bulk transformation still follows the
zero-temperature spinodal: for $f_s < 0.5$ the 5\,nK onsets lie within
1.6\,ms of the quantum-noise ones, with an advance that does not grow
systematically with $\tQ$, and give an exponent of $0.38 \pm 0.02$, against
$0.35 \pm 0.02$ with quantum noise ($\tQ = 50$--200\,ms; the 20\,ms onsets fall
after the end of the ramp at these thresholds). This small advance is what
larger seeds---the thermal occupation of the soft roton mode---produce by
shortening the growth after the instability. Thresholds near $f_s = 1$, in
contrast, are crossed 8--11\,ms after the zero-temperature spinodal at every
ramp time (after the spinodal, not inside the window), so that their advance
grows with $\tQ$ and their onset curve is nearly flat. A fixed ratio of seed
amplitudes in linear growth would advance all thresholds by similar amounts;
the additional early drop of $f_s$ for slow ramps can come from large thermal
fluctuations of the soft mode or from local crystalline regions forming just
after the spinodal, which our data do not separate. In holds above the window, at $90.8\,a_0$, the resolved roton
oscillations at 10\,nK agree with the quantum-noise frequencies, while the
5\,nK frequencies near the roton minimum cannot be determined
(Appendix~\ref{app:thermal}). Spinodal-controlled freeze-out of the bulk
transformation thus survives at 5\,nK for ramps up to at least 200\,ms,
provided the onset is detected with a threshold below the thermal
fluctuations of $f_s$. Deep droplets appear before the zero-temperature
spinodal in only 0 of 16, 1 of 16 and 2 of 32 realizations at $\tQ = 50$, 100
and 200\,ms. Preparation near the window produces a majority of early
nucleation events at the tested crossing rate; dominance of nucleation under
slower ramps from $92\,a_0$ remains an extrapolation.

These conclusions hold within the truncated-Wigner description, in which the
thermal fluctuations are a classical field and the temperature is not
controlled during the evolution. Scaling the noise cutoff by 0.7 and 1.3 leaves
the nucleation rate at $90.05\,a_0$ and 5\,nK unchanged within errors
($\Gamma = 0.15\pm0.04$ and $0.14\pm0.04\,\mu\mathrm{m}^{-1}\mathrm{s}^{-1}$,
against $0.11\pm0.03$; Appendix~\ref{app:noise}), even though the smaller
cutoff no longer seeds the roton modes directly. We return to these
limitations in Sec.~\ref{sec:discussion}.

\subsection{Low density: direct transition to a droplet crystal}\label{sec:lowdensity}

At $700\perum$ the first-order transition has a different character. The
modulated state consists of separated, self-bound droplets
($f_s \approx 0.1$ at the end of the ramps), and
near $\astar$ the optimal droplet spacing grows without bound. Forward ramps
from 87 to $83\,a_0$ again show no early transformation: the first droplets
appear at 84.38--84.46\,$a_0$ for slow ramps, just below the spinodal
$\arot = 84.506\,a_0$. The global superfluid fraction, however, falls only
about 10\,ms later, because the droplets form one by one and the crystal
spreads along the tube. This density-based diagnostic alone does not
establish an insulating phase, which would require correlations between
droplets beyond the mean-field eGPE description.

\begin{figure}[t]
\includegraphics[width=\columnwidth]{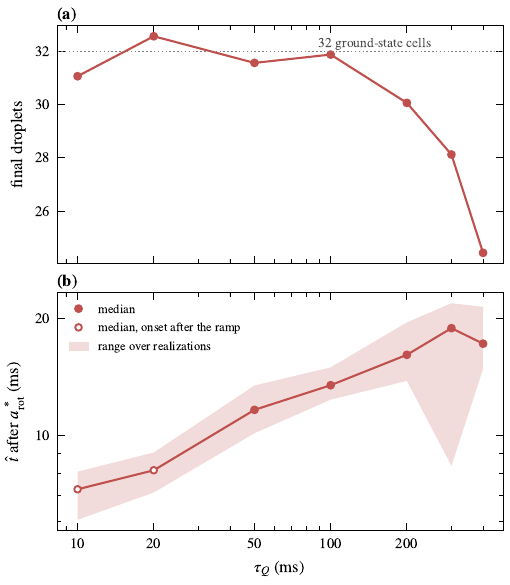}
\caption{Low density, $700\perum$, forward ramps from 87 to $83\,a_0$ with
quantum noise (16 realizations per $\tQ$). (a) Final number of droplets for a
tube that holds 32 ground-state cells: slower ramps select larger spacings.
(b) Freeze-out time after the spinodal, median and range over realizations.
Open circles mark onsets after the ramp endpoint, which are excluded from
the descriptive linear-ramp slope.}
\label{fig:droplets}
\end{figure}

Figure~\ref{fig:droplets} summarizes the low-density ramps. The median
freeze-out time measured from the spinodal increases from 7.3\,ms
($\tQ = 10$\,ms) to about 18\,ms (300\,ms) and then levels off; the curve is
not a single power law. The onsets at $\tQ=10$ and 20\,ms occur after the
ramp, during the final hold, and are shown as open circles. The remaining
50--400\,ms points have a descriptive effective slope of 0.22, with 16
realizations per $\tQ$; this is not a universal scaling exponent. The spread between realizations is
large: in one of sixteen realizations at $\tQ = 300$\,ms droplets formed
10\,ms earlier than in the others. Repeating the shorter ramps with other noise
seeds reproduces the same curve, shifted by only 0.2\,ms, so its shape is not a
statistical artefact.

The number of droplets formed is a useful observable here. For a tube that
holds 32 ground-state cells, fast ramps produce 31--33 droplets, while slower
ramps produce fewer: 30, 28 and 24 droplets for $\tQ = 200$, 300 and 400\,ms.
Slower ramps thus select larger droplet spacings, approaching the diverging
equilibrium spacing of this density.

On reverse ramps the droplet crystal does not melt as the supersolid does.
Raising $a_s$ from 83 to $87\,a_0$, the droplets first merge---the count falls
already below the uniform-state spinodal for slow ramps---and many survive the
whole ramp. Ramping further to $89\,a_0$, $3.7\,a_0$ above $\astar$, and holding
for 60\,ms, about a quarter of the droplets (7--8 of 32, counted with the strict
density criterion) remain. These are long-lived, self-bound droplets far
outside the equilibrium crystal phase, and they give the low-density
transition a much wider hysteresis than at high density.

\section{Discussion}\label{sec:discussion}

\textit{Why the spinodal controls the freeze-out.} A first-order transition
offers two routes from the uniform superfluid to the crystal: nucleation of the
new phase anywhere in the metastable window between $\astar$ and the spinodal,
or survival of the metastable state until the spinodal, where the uniform state
becomes linearly unstable. At zero temperature nucleation did not occur in our
simulations (no droplets during a 150\,ms hold inside the window), so the
system takes the second route. Close to the spinodal the dynamics is governed
by the softening roton mode, whose gap closes as the square root of the
distance to $\arot$, as at the continuous transition. This is why the
freeze-out time, measured from the spinodal, scales with an exponent close to
the mean-field value $1/3$, and why it is not a power law when measured from
$\astar$.

\textit{When nucleation wins.} At finite temperature the two routes compete.
The time spent inside the window grows linearly with $\tQ$, while the
probability of nucleation in that time grows with the nucleation rate, which
increases with temperature and with the distance below $\astar$. Fast ramps
reach the spinodal before a nucleus forms and show spinodal-controlled
freeze-out; longer residence in the metastable window can allow nucleation
to intervene. In our simulations the outcome is not set by the initial
temperature alone: the population of
the soft roton mode, which seeds nucleation, depends on how the system was
prepared and is not re-thermalized during the ramp
(Sec.~\ref{sec:nucleation}). At the tested crossing rate corresponding to
$\tQ=200$\,ms for the $92$--$88\,a_0$ ramp, changing the starting point
from $92$ to $90.5\,a_0$ raises the early-nucleation fraction from 2/32 to
11/16. The $\tQ\approx0.5$\,s value obtained from hold rates is a
protocol-dependent extrapolation; a crossover boundary has not been measured
for either preparation. A quantitative prediction of
$\tQ^{\times}(T)$ for experiments therefore requires a description in which the
thermal cloud keeps the soft mode at equilibrium, such as the stochastic
projected Gross--Pitaevskii equation~\cite{Blakie2008,Gardiner2003}. \Dy{}
experiments ramp over tens to hundreds of milliseconds, typically at
temperatures above the 2.5--10\,nK simulated here; since the nucleation rate
grows quickly with temperature, the nucleation route is likely to matter there,
and the spinodal regime is best reached with fast ramps and cold samples.
Varying the ramp rate at fixed temperature would locate the crossover.

\textit{Finite temperature and the reference point.} Two thermal effects
matter for an experimental test. First, thermal fluctuations lower the
superfluid fraction before the bulk transformation, so that onset thresholds
near $f_s = 1$ are crossed early and their onset curve is flattened; in our
simulations a threshold below the fluctuation level ($f_s < 0.5$) recovers
the spinodal scaling (Sec.~\ref{sec:nucleation}). Second, at the higher
temperatures and thermal fractions of experiments, thermal fluctuations soften
the roton in equilibrium and move the transition to larger
$a_s$~\cite{SanchezBaena2023,SanchezBaena2024}. We resolve no such softening of
the roton in our classical field at 10\,nK (Appendix~\ref{app:thermal}), which
populates only modes below $2k_BT$ and adds less than 1\% of the atoms.
Because a reference shift of $0.03\,a_0$ already changes the fitted exponent
by about 0.02, the spinodal used to analyze an experiment should be computed
with a finite-temperature theory or measured independently, for instance from
the softening of the roton.

\textit{Low density.} At $700\perum$ the transition creates self-bound
droplets. Their number depends on the ramp rate, and they survive far above the
transition on reverse ramps. Both effects make the low-density side a natural
place to study defect formation and metastability with in-situ droplet
counting.

\textit{Limitations.} (i) The truncated-Wigner approximation treats
fluctuations as a classical field: it captures the seeding of instabilities and
thermally seeded nucleation, but there is no bath: the
population of the soft roton mode is set by the preparation and by adiabatic
following rather than by the temperature, so nucleation rates depend on the
protocol (Sec.~\ref{sec:nucleation}), although they are insensitive to a
$\pm30\%$ change of the noise cutoff (Appendix~\ref{app:noise}). Quantitative
rates require a thermalizing c-field method. (ii) We simulate an infinite,
periodic tube; experiments use finite traps, where the density varies along
the axis and the transition is crossed at different times in different places.
(iii) Most parameter sets use 8--24 realizations; the exponents carry bootstrap
and systematic errors, but rare events (early transformations, nucleation in
ramps) are counted with few realizations. (iv) Our transition points come from
static ground states and Bogoliubov spectra on the numerical grid of the
dynamics at zero temperature; the freeze-out exponent is sensitive to them,
and at finite temperature the onset also depends on the detection threshold
(Sec.~\ref{sec:nucleation}). (v) The transition point $\astar$ refers to
states of uniform density; in the thermodynamic limit, phase separation into
coexisting regions replaces it by a narrow coexistence window
(Appendix~\ref{app:coexistence}).

\textit{Conclusion.} At zero temperature, the first-order
superfluid--supersolid transition of a dipolar gas in a tube is crossed through
its spinodal, with a KZ-like freeze-out whose exponent is close to that of the
continuous transition over the tested linear-ramp range. Initial thermal
fluctuations advance the first drop of the superfluid fraction, while the
bulk transformation still follows the spinodal, and they open a nucleation
route inside the metastable window within the truncated-Wigner dynamics. Its
relative importance at a fixed crossing rate
depends on how the soft roton mode was prepared. Hold rates motivate a
crossover for slower ramps, but do not quantitatively locate it for the
production ramp protocol. These results identify preparation, ramp rate and
fluctuation strength as controls of crystallization that experiments can
probe through hysteresis and droplet counting.

\begin{acknowledgments}
The software, the numerical experiments, the analysis and the draft of this
manuscript were developed with substantial assistance from a generative AI
coding agent, Claude (Anthropic; Claude Code with the model Claude Opus 5.5),
under the author's direction; once, OpenAI Codex (model gpt-6-astra) was used
for a read-only review of part of the code. AI-generated code and results were
checked by a regression suite of 126 tests against analytic and independent
numerical results, by the reproduction of published results, by convergence
studies, and by provenance records for every run. The author reviewed the code,
results and text and takes full responsibility for the content.
\end{acknowledgments}

\section*{Data availability}
The simulation code (DipGPE), the run records (command, parameters, code version
and environment of every run) and the analysis scripts that produce every
number and figure in this paper, including the coexistence, localized-state and
thermal analyses of Appendices E and F, are available at
\url{https://github.com/kuchris/spinodal-freezeout-dipolar-supersolid}.
The release used here is archived in Zenodo~\cite{WongCode2026} under
\href{https://doi.org/10.5281/zenodo.23181065}{doi:10.5281/zenodo.23181065}
(all versions). The reduced data---the per-realization observables of every
simulation and the static results used here, with their parameters and
provenance---are archived in Zenodo~\cite{WongData2026} under
\href{https://doi.org/10.5281/zenodo.23181218}{doi:10.5281/zenodo.23181218}
(all versions).

\appendix

\section{Full 3D model versus the quasi-1D reduced model}\label{app:quasi1d}

Earlier studies of the tube~\cite{Blakie2020,Kirkby2025} used a reduced model
in which the transverse wavefunction is a fixed or variational Gaussian with
width $w$ and ellipticity $\eta$. Our implementation of this model reproduces
both references when given their ingredients: the roton instability at 81.221
and $88.578\,a_0$ for 700 and $6250\perum$ (Blakie \textit{et al.}: 81.2,
88.6), the variational widths at $88\,a_0$ ($w = 1.078\,\mu$m, $\eta = 4.248$;
Kirkby \textit{et al.}: 1.08\,$\mu$m, 4.25), and the critical point and
freeze-out exponent of Ref.~\cite{Kirkby2025} ($a_c = 91.08\,a_0$ versus
91.05; $\zeta = 0.355\pm0.004$ versus 0.346--0.352).

Two approximations in those ingredients matter. First, the closed-form
expression for the quasi-1D dipolar kernel is exact only for a round cross
section ($\eta = 1$); for $\eta = 4.25$ it deviates by up to about 2\% from the
exact kernel at intermediate wavenumbers, which we compute by a
one-dimensional quadrature and verify against a direct transverse integral (to
$10^{-12}$) and against the 3D code (to $10^{-11}$ in the dipolar energy). With
fixed widths this shifts the roton instability from 91.08 to $90.54\,a_0$.
Second, the LHY term in Ref.~\cite{Blakie2020} uses
$Q_5 \approx 1 + 3\epsilon_{\mathrm{dd}}^2/2$, 5--6\% below the exact value here. With both
corrected, the reduced model lies 2.3--4.5\,$a_0$ below the full 3D eGPE
($\arot = 80.26$, 90.03, $85.32\,a_0$ versus 84.51, 92.31, $89.85\,a_0$ at 700,
2500, $6250\perum$), further than with both approximations
(0.8--3.3\,$a_0$): the approximations partly cancel the error of the Gaussian
ansatz. The reduced model also gives a first-order window $\astar-\arot$ about
2.5 times wider than the full model (2.1 and $1.0\,a_0$ versus 0.77 and
$0.41\,a_0$). We therefore use it only as a cross-check.

\section{Convergence}\label{app:convergence}

The available checks are at $6250\perum$, forward ramps, $\tQ = 50$\,ms unless stated
(Table~\ref{tab:convergence}).
Because the stable time step scales with the square of the grid spacing,
the grid comparison also changes the time step; the time step is therefore
tested separately by halving it (10 to 5\,$\mu$s) at fixed grid for
$\tQ = 20$ and 50\,ms with the same noise realizations. The median onset
times are unchanged at all thresholds ($f_{\mathrm{th}} = 0.98$, 0.9, 0.7),
and no realization moves by more than one record (0.1\,ms).

\begin{table}[!htbp]
\caption{Convergence of the median freeze-out time at $6250\perum$,
$\tQ = 50$\,ms; indented rows repeat the check at another $\tQ$.}
\label{tab:convergence}
\setlength{\tabcolsep}{2pt}
\renewcommand{\arraystretch}{1.12}
\begin{ruledtabular}
\begin{tabular}{@{}lll@{}}
\tcell{0.31}{change} & \tcell{0.30}{freeze-out $\that$} & \tcell{0.33}{effect} \\
\hline
\tcell{0.31}{tube length 32 $\to$ 64 cells} & \tcell{0.30}{12.16 $\to$ 12.26\,ms} & \tcell{0.33}{$<1\%$} \\
\tcell{0.31}{grid spacing\newline 0.25 $\to$ 0.1875\,$l$} & \tcell{0.30}{12.46 $\to$ 12.26\,ms} & \tcell{0.33}{$\le 1.6\%$;\newline \mbox{$\Delta\zeta = +0.015$}} \\
\tcell{0.31}{\quad $\tQ = 100$\,ms} & \tcell{0.30}{15.93 $\to$ 15.83\,ms} &  \\
\tcell{0.31}{transverse box 24 $\to$ 32\,$l$} & \tcell{0.30}{12.46 $\to$ 12.22\,ms} & \tcell{0.33}{2\%;\newline \mbox{$\Delta\arot = -0.003\,a_0$}} \\
\tcell{0.31}{complex64 $\to$\newline complex128} & \tcell{0.30}{identical per realization} & \tcell{0.33}{none} \\
\tcell{0.31}{time step\newline 10 $\to$ 5\,$\mu$s} & \tcell{0.30}{12.46 $\to$ 12.46\,ms} & \tcell{0.33}{none ($\le 0.1$\,ms per realization)} \\
\tcell{0.31}{\quad $\tQ = 20$\,ms} & \tcell{0.30}{8.93 $\to$ 8.93\,ms} &  \\
\tcell{0.31}{noise seeds ($700\perum$, all $\tQ$)} & \tcell{0.30}{$+0.2$\,ms} & \tcell{0.33}{shape unchanged} \\
\end{tabular}
\end{ruledtabular}
\end{table}

\section{Noise sampling}\label{app:noise}

Each mode $u_m$ of the initial state receives a complex Gaussian amplitude
$a_m$ with $\langle|a_m|^2\rangle = 1/(e^{\varepsilon_m/k_BT}-1) + 1/2$. Modes are
plane waves along the tube, $e^{ikz}/\sqrt{L}$ with $k = 2\pi j/L$, $j\neq0$,
times products of transverse harmonic-oscillator eigenfunctions, with energies
$\varepsilon = \hbar^2k^2/2m + \hbar\omega_\perp(n_x+n_y)$; modes with
$\varepsilon < 2k_BT$ are kept. For quantum noise only the half quantum is used,
in the transverse ground state and for $|k|$ below the inverse healing length
$1/\xi = \sqrt{m\mu}/\hbar$ ($\mu$ measured from the transverse zero-point
energy). The thermal prescription of Ref.~\cite{Kirkby2025}, as printed,
cannot be reproduced exactly: with standard free-particle energies it adds
about 15\% of the atoms at 20\,nK rather than the stated 3--5\%; we therefore
anchor to their quantum-noise results and use temperatures of 2.5--10\,nK, at
which our thermal noise adds 0.2--0.8\% of the atoms.

\textit{Cutoff sensitivity.} Scaling the quantum cutoff $k_{\max} = 1/\xi$ by
0.7 and 1.3 (16 realizations, $6250\perum$, forward ramps) changes the median
freeze-out time by at most 0.4\,ms (2.5\%), less than the spread between
realizations (Table~\ref{tab:cutoff}). The variation of $\zeta$ ($\le 0.02$,
without a trend) is within the quoted error; this is expected because the
growth time from a seed to a crystal depends only logarithmically on the seed
amplitude. Scaling the thermal energy cutoff $2k_BT$ by 0.7 and 1.3 (holds at
$90.05\,a_0$, 5\,nK, 16 realizations) gives $\Gamma = 0.15\pm0.04$ and
$0.14\pm0.04\,\mu\mathrm{m}^{-1}\mathrm{s}^{-1}$, against $0.11\pm0.03$ for the
standard cutoff. With the cutoff reduced to $1.4\,k_BT$ the roton modes
($\hbar^2k_{\mathrm{rot}}^2/2m \approx 1.6\,k_BT$) are not seeded directly and the
roton-band power at the start of the hold halves, yet the rate is unchanged: in
the holds the seeds of nucleation are generated dynamically from the other
thermal modes.

\begin{table}[!htbp]
\caption{Median freeze-out time for a scaled quantum-noise cutoff.}
\label{tab:cutoff}
\begin{ruledtabular}
\begin{tabular}{cccc}
$k_{\max}$ & $\that$ ($\tQ = 20$\,ms) & $\that$ ($\tQ = 100$\,ms) & two-point $\zeta$ \\
\hline
$0.7/\xi$ & 9.03\,ms & 15.52\,ms & 0.337 \\
$1/\xi$   & 8.93\,ms & 15.92\,ms & 0.360 \\
$1.3/\xi$ & 9.13\,ms & 15.82\,ms & 0.342 \\
\end{tabular}
\end{ruledtabular}
\end{table}

\section{Fit procedure and error budget}\label{app:errors}

The linear-ramp fits exclude ramp times whose median $f_s=0.98$ onset occurs
after the end of the ramp. Medians use the upper middle realization for an
even sample count, and bootstrap errors use 2000 resamplings. Threshold checks
use a fixed, shorter ramp window in which the median onset is before the
endpoint for every threshold, $f_{\mathrm{th}}=0.90$, 0.95, 0.98 and 0.99.
The difference between the primary and common-window slopes is included
as a fit-window sensitivity.

\begin{table}[!htbp]
\caption{Exponent sensitivity components. Columns are bootstrap standard
deviation, threshold half-range, reference shift, fit-window shift, and
combined estimate. Every combined estimate also includes a 0.015
spatial-sensitivity allowance. The low-density row is descriptive.}
\label{tab:errors}
\begin{ruledtabular}
\begin{tabular}{lccccc}
ensemble & boot. & thresh. & ref. & window & combined \\
\hline
6250 forward & 0.0074 & 0.0100 & 0.0080 & 0.0010 & 0.0211 \\
2500 forward & 0.0064 & 0.0044 & 0.0047 & 0.0061 & 0.0186 \\
700 forward & 0.0260 & 0.0131 & 0.0209 & 0.0000 & 0.0389 \\
6250 reverse & 0.0016 & 0.0096 & 0.0168 & 0.0000 & 0.0245 \\
\end{tabular}
\end{ruledtabular}
\end{table}

The 0.015 spatial allowance comes from the recorded 6250-forward grid
comparison and is carried as a proxy for the other ensembles; it is not a
separately measured grid effect in each case. These systematic sensitivities
are not probabilistic confidence intervals. Onset locations in
Fig.~\ref{fig:hysteresis}(a) use the actual piecewise ramp-and-hold
schedule.

\section{Coexistence and localized states}\label{app:coexistence}

\begin{figure*}[t]
\includegraphics[width=\textwidth]{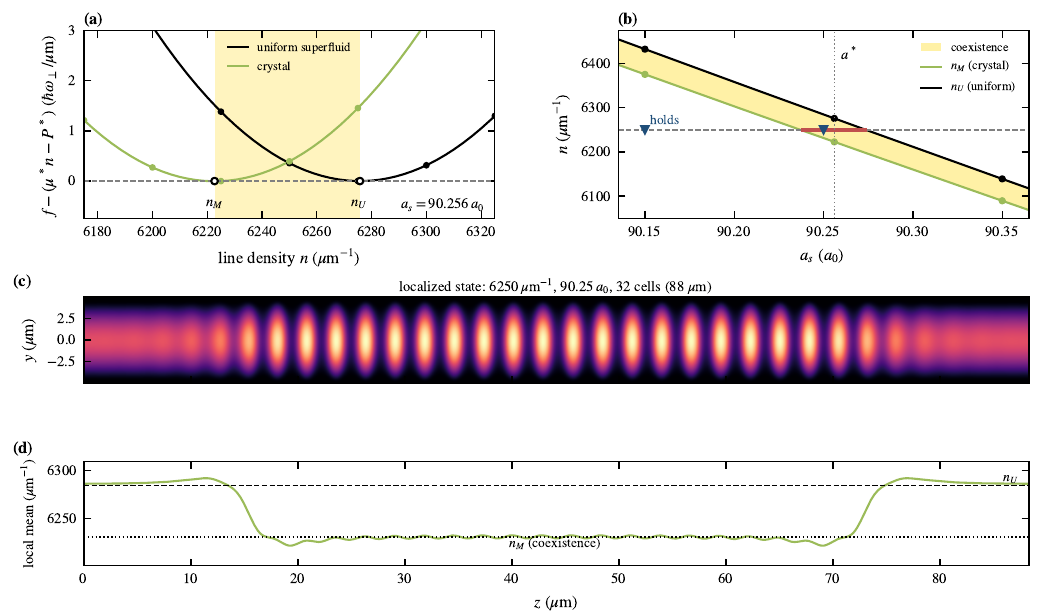}
\caption{Coexistence and localized states at $6250\perum$. (a) Energy per
length of the uniform and crystal branches at $a_s = 90.256\,a_0$ minus the
common tangent $\mu^* n - P^*$; the tangent touches the branches at the
coexisting densities $n_M$ and $n_U$ (shaded). (b) Coexisting densities versus
$a_s$ (points: Maxwell construction; lines: quadratic fits); at
$n = 6250\perum$ (dashed) phase separation is favorable in the window marked
in red, which contains $\astar$ and the hold at $90.25\,a_0$ (triangles:
holds of Sec.~\ref{sec:nucleation}). (c) Column density $\int dx\,|\Psi|^2$ of
the lowest localized state in the 32-cell tube at $90.25\,a_0$ and (d) its line
density smoothed over about one cell, with the coexisting densities.}
\label{fig:coexistence}
\end{figure*}

\textit{Maxwell construction.} At fixed $a_s$ we compute, for linear densities
$n$ in steps of $25\perum$ ($10\perum$ at $2500\perum$), the $z$-uniform ground state and the crystal ground
state in one cell, the latter at three cell lengths with the energy minimized
by a parabola in the cell length; each crystal state is continued from the
previous density, so that the branch is followed also where it is metastable.
From the energy per length $f(n) = nE/N$ and the chemical potential $\mu$ of
each state follows the one-dimensional pressure $P = \mu n - f$. Two phases
coexist where their branches cross in the $(\mu, P)$ plane; the coexisting
densities are $n = \partial P/\partial\mu$ at the crossing. Equivalently, the
common tangent to $f(n)$ touches the crystal branch at $n_M$ and the uniform
branch at $n_U$ [Fig.~\ref{fig:coexistence}(a)]. The thermodynamic identities
$\mu = \partial f/\partial n$ and $\partial P/\partial\mu = n$ hold along both
branches with relative finite-difference residuals below $2\times10^{-5}$, and
the window does not change when the cell length is
fixed instead of optimized (widths 0.0366--0.0369\,$a_0$ for 4.2, 4.3 and
$4.4\,l$; the position moves by at most $0.011\,a_0$). Refining the density
steps to $10\perum$ or the grid spacing to $0.1875\,l$ changes the window by
less than $10^{-5}\,a_0$, and enlarging the transverse box to $32\,l$ shifts it
by $-0.002\,a_0$ and widens it by less than 1\%. The uniform-density crossing
reproduces $\astar = 90.2555\,a_0$.

At $6250\perum$ the coexisting densities are $n_M = 6375$, 6223 and
$6089\perum$ and $n_U = 6432$, 6276 and $6139\perum$ at $a_s = 90.15$,
90.256 and $90.35\,a_0$; they fall by about $1450\perum$ per $a_0$ along the
transition line [Fig.~\ref{fig:coexistence}(b)]. A tube at $n = 6250\perum$
is therefore entirely crystalline below $90.237\,a_0$, entirely superfluid
above $90.274\,a_0$, and phase separated in between, where the length fraction
of the crystal follows the lever rule $(n_U - n)/(n_U - n_M)$ (0.65 at
$90.25\,a_0$). In the thermodynamic limit, phase separation lowers the energy
near $\astar$ by about $6\times10^{-5}\,\hbar\omega_\perp$ per atom relative to
the better homogeneous state ($5.7\times10^{-5}$ at $90.256\,a_0$).

At $2500\perum$ the crystal branch emerges continuously from the uniform
branch at $a_s = 92.25$ and $92.20\,a_0$ (onsets near 2360 and $2280\perum$).
Close to the onset $f_M - f_U = -(\kappa/2)(n - n_c)^2$, so that the slope
$\partial\mu_M/\partial n = \partial\mu_U/\partial n - \kappa$ of the crystal
branch (its inverse compressibility) would turn negative, and the states would
coexist, if $\kappa$ exceeded $\partial\mu_U/\partial n$~\cite{Steinberg2025}.
We find $\kappa/(\partial\mu_U/\partial n) = 0.004$--0.009 for the three cell
lengths, and no crossing of the two branches in the $(\mu, P)$ plane between
2200 and $2700\perum$: the transition at this density is continuous also when
the density may vary along the tube.

\textit{Localized states.} In a tube of 32 cells ($88\,\mu$m, the tube of the
dynamics) and 64 cells we start from a crystal patch at $n_M$ next to uniform
superfluid at $n_U$ and minimize the energy at fixed atom number
($n = 6250\perum$). The minimization converges to stationary localized
states [Fig.~\ref{fig:coexistence}(c)] whose crystal and superfluid regions
have the coexisting densities to within $6\perum$ (0.1\%)
[Fig.~\ref{fig:coexistence}(d)]; their interfaces are about $9\,\mu$m wide.
The lowest 32-cell state at $90.25\,a_0$ is a local minimum: after a random
perturbation of relative amplitude $10^{-3}$ the minimization returns to the
same energy (to $10^{-14}$).
Different initial patches lead to localized states that differ by one
droplet and by $7\times10^{-8}\,\hbar\omega_\perp$ per atom, consistent with
the several branches of localized states found in Ref.~\cite{Steinberg2025}.
These are local energy minima, not the unstable critical nuclei that set the
nucleation barrier of Sec.~\ref{sec:nucleation}; finding those requires a
search for saddle points, which we leave for future work.
Subtracting the bulk energy of the coexisting phases gives an interface energy
$\sigma = 47.3$--$47.8\,\hbar\omega_\perp$ ($\approx 340\,\mathrm{nK}\,k_B$);
comparing the energies of the 32- and 64-cell tubes, which does not use the
bulk reference, gives 47.9 and $47.5\,\hbar\omega_\perp$ at 90.25 and
$90.256\,a_0$. At $90.25\,a_0$ the localized state lies $1.4\times10^{-4}$
(32 cells) and $5.7\times10^{-5}\,\hbar\omega_\perp$ per atom (64 cells) above
the homogeneous crystal. Phase separation pays off only when its bulk gain
exceeds $2\sigma$, i.e., in tubes longer than about $250\,\mu$m near $\astar$
($265\,\mu$m at $90.256\,a_0$) and $513\,\mu$m at $90.25\,a_0$. The computed
localized states in the 88\,$\mu$m tube lie above the homogeneous minimum, and
the bulk-plus-interface estimate predicts the same throughout the window. The
thermal holds of Sec.~\ref{sec:nucleation} contain 0.4\% more atoms, which
moves the window by about $-0.02\,a_0$ ($90.25\,a_0$ stays inside it), and
they are not at zero temperature, so this static comparison applies to them
only approximately.

\section{Thermal effects on the onset}\label{app:thermal}

\textit{Ramps.} Table~\ref{tab:thermal} compares the median onset times after
the zero-temperature spinodal for forward ramps from 92 to $88\,a_0$ at
$6250\perum$ with quantum noise (16 realizations) and with thermal noise at
5\,nK (16--24 realizations), for three thresholds. A rigid shift of the onset
curve would advance all thresholds equally; instead the advance falls from
11.3 to 0.9\,ms between $f_s < 0.98$ and $f_s < 0.5$ at $\tQ = 200$\,ms. As in
Appendix~\ref{app:errors}, medians use the upper middle realization for an
even sample count, bootstrap errors use 2000 resamplings, and onsets after the
end of the linear ramp are excluded from the exponents.

\begin{table}[!htbp]
\caption{Advance (ms) of the median onset with thermal noise at 5\,nK relative
to quantum noise, for three thresholds, and the exponent $\zeta$ of the median
onset after the zero-temperature spinodal versus $\tQ$. Asterisks mark onsets after
the end of the ramp, which are excluded from the exponents, so the exponents are
fitted over $\tQ = 20$--200\,ms for $f_s < 0.98$ and 50--200\,ms for
$f_s < 0.7$ and 0.5. All uncertainties are bootstrap standard deviations
only; the combined estimate for the quantum-noise exponent at $f_s < 0.98$ is
$\pm0.021$ (Table~\ref{tab:errors}).}
\label{tab:thermal}
\setlength{\tabcolsep}{3pt}
\begin{ruledtabular}
\begin{tabular}{cccc}
$\tQ$ (ms) & $f_s < 0.98$ & $f_s < 0.7$ & $f_s < 0.5$ \\
\hline
20  & $0.70\pm0.24$ & $0.60\pm0.20^*$ & $0.50\pm0.23^*$ \\
50  & $1.90\pm0.78$ & $1.60\pm0.50$ & $1.20\pm0.32$ \\
100 & $6.50\pm1.88$ & $3.00\pm1.22$ & $1.60\pm0.40$ \\
200 & $11.30\pm1.77$ & $7.40\pm1.70$ & $0.90\pm0.50$ \\
\hline
$\zeta$, quantum & $0.363\pm0.007$ & $0.345\pm0.014$ & $0.347\pm0.015$ \\
$\zeta$, 5\,nK & $0.04\pm0.07$ & $0.17\pm0.08$ & $0.38\pm0.02$ \\
\end{tabular}
\end{ruledtabular}
\end{table}

\textit{Roton frequency above the window.} To test whether thermal noise
softens the roton away from the spinodal, we hold the gas at $90.8\,a_0$, above
the end of the crystal branch, after a 20\,ms ramp from $92\,a_0$, and record
the line density of every realization for 250\,ms (8 realizations at 5 and
10\,nK, 4 with quantum noise as the zero-temperature reference; the noise adds
0.40, 0.84 and 0.01\% of the atoms). For each tube mode $k_m = 2\pi m/L$ the
realization-averaged autocorrelation of its Fourier amplitude over
40--280\,ms is fitted by $A e^{-\gamma\tau}\cos\omega\tau$ for lags up to 60,
100 and 150\,ms, with bootstrap errors. With quantum noise the roton minimum is
at $m = 31$ (wavelength $2.85\,\mu$m), $\omega_{\mathrm{rot}} =
0.401\,\omega_\perp$; with $\omega_{\mathrm{rot}}^2 \propto a_s - \arot$, a
change of $\omega^2$ corresponds to an effective shift
$-\delta(\omega^2)(90.8 - 89.845)\,a_0/\omega_{\mathrm{rot}}^2$ of the
instability. At 10\,nK the three modes at the minimum are weakly damped
($\gamma \le 0.02\,\omega_\perp$); compared mode by mode with the quantum-noise
frequencies they give effective shifts of $0.001 \pm 0.007$,
$0.002 \pm 0.007$ and $0.000 \pm 0.006\,a_0$ for the three lag windows, i.e., no
resolved softening. At 5\,nK the modes at and just above the minimum are
strongly damped ($\gamma \approx 0.1$--$0.15\,\omega_\perp$); the single
remaining fit ($m = 31$) gives $0.005 \pm 0.16\,a_0$ and is inconclusive. Below
the minimum the weakly damped modes lie 0.7--1.8\% (5\,nK) and 0.3--1.2\%
(10\,nK) below their quantum-noise frequencies. These nonequilibrium fits test
only the classical field at these temperatures; they do not address the
equilibrium shift of Refs.~\cite{SanchezBaena2023,SanchezBaena2024} at the
higher temperatures of experiments.

\bibliography{refs}

\end{document}